\documentclass[preprint]{aastex631}

\newcommand{\kms}{km\,s$^{-1}$}

\graphicspath{{./}{figures/}}

\received{January 1, 2020}
\revised{January 7, 2020}
\accepted{\today}

\submitjournal{ApJ}

\shorttitle{APOGEE Line List}
\shortauthors{Smith et al.}

\begin{document}

\title{The APOGEE Data Release 16 Spectral Line List}


\correspondingauthor{Verne V. Smith}
\email{vsmith@noao.edu}

\author[0000-0002-0134-2024]{Verne V. Smith}
\affiliation{NSF's NOIRLab, 950 N. Cherry Ave. 
Tucson, AZ 85719 USA}

\author[0000-0002-3601-133X]{Dmitry Bizyaev}
\affiliation{Apache Point Observatory and New Mexico State University, Sunspot, NM 88349}
\affiliation{Sternberg Astronomical Institute, Moscow State University, Moscow, 119992, Russia}

\author[0000-0001-6476-0576]{Katia Cunha}
\affiliation{Steward Observatory, 
Department of Astronomy \\
University of Arizona, Tucson,  AZ 85721 USA}
\affiliation{Observat\'orio Nacional, MCTI, Rio de Janeiro, Brazil}

\author[0000-0003-0509-2656]{Matthew D. Shetrone}
\affiliation{McDonald Observatory and Department of Astronomy \\
University of Texas, Austin, TX 78712}

\author[0000-0002-7883-5425]{Diogo Souto}
\affiliation{Universidade Federal de Sergipe, Brazil}

\author[0000-0002-0084-572X]{Carlos Allende Prieto}
\affiliation{Instituto de Astrof\'isica de
Canarias (IAC), E-38205 La Laguna, Tenerife, Spain}
\affiliation{Universidad de La Laguna (ULL), Departamento de Astrof\'isica,
38206 La Laguna, Tenerife, Spain}

\author[0000-0002-6939-0831]{Thomas Masseron}
\affiliation{Instituto de Astrof\'isica de
Canarias (IAC), E-38205 La Laguna, Tenerife, Spain}
\affiliation{Universidad de La Laguna (ULL), Departamento de Astrof\'isica,
38206 La Laguna, Tenerife, Spain}

\author{Szabolcs M\'esz\'aros}
\affiliation{ELTE E\"otv\"os Lor\'and University, Gothard Astrophysical Observatory, 9700 Szombathely, Szent Imre H. st. 112, Hungary}
\affiliation{MTA-ELTE Exoplanet Research Group}

\author[0000-0002-4912-8609]{Henrik Jonsson}
\affiliation{Materials Science and Applied Mathematics, Malm\"o University, SE-205 06 Malm\"o, Sweden}
\affiliation{Lund Observatory, Department of Astronomy and Theoretical Physics, Lund University, Box 43, SE-22100 Lund, Sweden}

\author[0000-0001-5388-0994]{Sten Hasselquist}
\affiliation{Department of Physics \& Astronomy, University of Utah, Salt Lake City, UT 84112}

\author[0000-0001-5832-6933]{Yeisson Osorio}
\affiliation{Instituto de Astrofisica de
Canarias (IAC), E-38205 La Laguna, Tenerife, Spain}
\affiliation{Universidad de La Laguna (ULL), Departamento de Astrofisica,
38206 La Laguna, Tenerife, Spain}

\author[0000-0002-1693-2721]{D. A. Garc\'ia-Hern\'andez}
\affiliation{Instituto de Astrof\'isica de
Canarias (IAC), E-38205 La Laguna, Tenerife, Spain}
\affiliation{Universidad de La Laguna (ULL), Departamento de Astrof\'isica,
38206 La Laguna, Tenerife, Spain}

\author[0000-0002-0398-4434]{Bertrand Plez}
\affiliation{LUPM, UMR 5299, Universite de Montpellier, CNRS, 34095 Montpellier, France}

\author[0000-0002-1691-8217]{Rachael L. Beaton}
\affiliation{Department of Astrophysical Sciences, Princeton University, 4 Ivy Lane, Princeton, NJ  08544}

\author[0000-0002-9771-9622]{Jon Holtzman}
\affiliation{Astronomy Department, New Mexico State University, Las Cruces, NM 88003}

\author[0000-0003-2025-3147]{Steven R. Majewski}
\affiliation{Department of Astronomy, University of Virginia, PO Box 400325, Charlottesville, VA 22904}

\author[0000-0003-1479-3059]{Guy S. Stringfellow}
\affiliation{Center for Astrophysics and Space Astronomy, Department of Astrophysical and Planetary Sciences, University of Colorado, Boulder, CO 80309}

\author{Jennifer Sobeck}
\affiliation{Department of Astronomy, Box 351580, University of Washington, Seattle, WA 98195}


\begin{abstract}

The updated $H$-band spectral line list (from $\lambda$15,000 - 17,000\AA)
adopted by the Apache Point Observatory Galactic Evolution Experiment
(APOGEE) for the SDSS IV Data Release 16 (DR16) is presented here.  The
APOGEE line list is a combination of atomic and molecular lines with data
from laboratory, theoretical, and astrophysical sources.  Oscillator
strengths and damping constants are adjusted using high signal-to-noise,
high-resolution spectra of the Sun and $\alpha$ Boo (Arcturus) as ``standard
stars''.  Updates to the DR16 line list, when compared to the previous DR14
version, are the inclusion of molecular H$_{2}$O and FeH lines, as well as a
much larger (by a factor of $\sim$4) atomic line list, which includes
significantly more transitions with hyperfine splitting.  More recent
references and line lists for the crucial molecules CO and OH were used, as
well as for C$_{2}$ and SiH.  In contrast to DR14, DR16 contains measurable
lines from the heavy neutron-capture elements cerium (as Ce II), neodymium
(as Nd II), and ytterbium (as Yb II), as well as one line from rubidium (as
Rb I), that may be detectable in a small fraction of APOGEE red giants.

\end{abstract}



\section{Introduction}

The Apache Point Observatory Galactic Evolution Experiment (APOGEE; Majewski et al. 2017) is one program within the Sloan Digital Sky Surveys (SDSS), as part of SDSS-III (Eisenstein et al. 2011) and SDSS-IV (as APOGEE-2; Blanton et al. 2017).  APOGEE\footnote{In this paper we will refer to  both APOGEE and APOGEE-2 collectively as ``APOGEE''.}
is a large spectroscopic survey expected to exceed $\sim$700,000 stars, consisting primarily of Galactic red giants from all stellar populations, but also including the Magellanic Clouds, and other nearby dwarf galaxy red giants, as well as significant numbers of cool (FGKM) dwarf stars (Zasowski et al. 2013; 2017). 
The data from APOGEE consist of high-resolution ($R\sim22,500$), near-infrared (NIR) spectra covering the wavelength range $\lambda$1.51 $-$ 1.70$\mu$m; the spectra are obtained from two, 300-object, fiber-fed spectrographs (Wilson et al. 2019), one in the North mated to the SDSS 2.5m telescope (Gunn et al. 2006) at Apache Point Observatory (APO), with the added capability to collect some spectra via a second fiber-feed to the New Mexico State University 1m telescope (Holtzman et al. 2010), and a second spectrograph in the South on the du Pont 2.5m telescope (Bowen \& Vaughan 1973) at Las Campanas Observatory (LCO).

The data are reduced to sky-subtracted, one-dimensional spectra via the pipeline described in Nidever et al. (2015).  Fundamental stellar parameters (effective temperature, $T_{\rm eff}$, surface gravity, $\log{g}$, and overall metallicity, [M/H]), as well as detailed chemical abundances, are derived from the stellar spectra using the APOGEE Stellar Parameter and Chemical Abundance Pipeline (ASPCAP; Garc\'ia Pe\'erez et al. 2016).  ASPCAP finds the best matches between the observed spectra and an extensive library of synthetic spectra; these model spectra are generated using a detailed spectral line list.  New APOGEE results are included in the recent SDSS data release, DR16 (Ahumada et al. 2019), with J\"onsson et al. (2020) providing a detailed overview, description, and discussion of the DR16 APOGEE results.  As part of DR16, the APOGEE team carried out a significant update to the line list used in the previous release, DR14, that was described in Shetrone et al. (2015).  The updates to the DR16 APOGEE line list are presented and described here.  

\section{Line List Data Sources}

The APOGEE line list consists of both atomic and molecular lines; this section describes the sources and references for the spectral lines that comprise the DR16 line list that were used to compute the stellar spectral libraries.  When discussing the transition oscillator strengths, $f$, the product of the transition
statistical weight, $g$, and $f$ will typically be quoted, usually as the value $\log(gf)$.  The APOGEE line list is restricted to a wavelength range of $\lambda$15,000 -- 17,000\AA, only slightly larger than the wavelengths covered by the APOGEE detectors.

\subsection{Atomic Lines}

A base atomic line list was created using a recent Kurucz (2017) line list from \url{http://kurucz.harvard.edu/linelists/gfnew/gfallwn08oct17.dat}, which includes large numbers of transitions split by hyperfine splitting (hfs).  The previous APOGEE line list from DR14 used, as a starting point, an earlier version of a Kurucz list that contained a smaller number of lines, by about a factor of 4.  Figure 1 illustrates the differences in the number of atomic lines from the DR14 list (bottom panel) when compared to the DR16 list (top panel), where the number of lines is shown for each element (plotted as atomic number, $Z$).  
The large increase in the number of spectral lines for all elements is apparent, with particularly large numbers of hyperfine-split line components for V and Co in the new list.  There are also a significant number of new heavy-element lines from Sr to Pd ($Z=38$ -- 46) in the DR16 list.
The DR16 line list also contains a few heavy elements that were not included in the DR14 list, including Ce II (from Cunha et al. 2017), Nd II (from Hasselquist et al. 2017), and Yb II (discussed here).

It should be noted, however, that the atomic line list for APOGEE contains many more lines than are typically detectable in $H$-band stellar spectra of cool giants, which are the main targets of the APOGEE survey, or, for example, M-dwarfs, which have also become a significant component of APOGEE; all lines in the Kurucz (2017) line list have been retained in the APOGEE line list as these may aid future investigations of, for example, very hot stars and nebular features, or help guide future laboratory efforts.
Literature and web sources for the all-important $gf$-values were also reviewed and updated as needed and these are discussed in the next section.

\begin{figure}
\epsscale{0.90}
\plotone{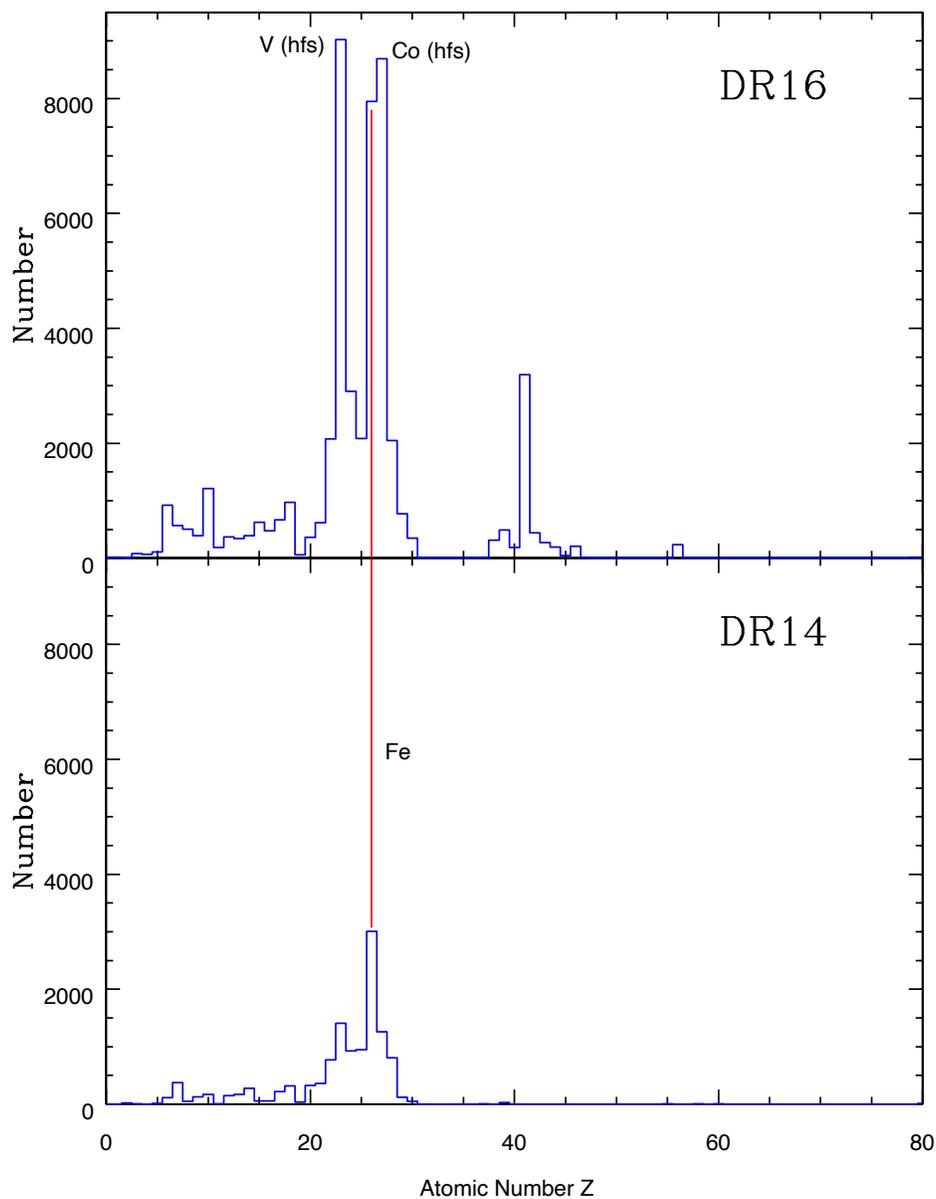}
\caption{Comparison of the numbers of atomic lines as a function of element (atomic number, $Z$) for the DR14 base atomic line list (bottom panel) and the DR16 base atomic line list (top panel).  The position of Fe ($Z=26$) is indicated in both panels by the red line.  The starting atomic line list for DR16 contains more than $4\times$ the number of lines from DR14, with a large increase in transitions with hfs, as well as a significant number of new heavy-element ($Z>30$) lines. The same vertical scale is used for both panels to highlight the increase in numbers of atomic transitions in DR16 compared to DR14.
\label{f1}}
\end{figure}

\subsubsection{Oscillator Strengths}

The APOGEE abundance pipeline, ASPCAP, uses pre-defined spectral windows to fit synthetic and model spectra and derive the chemical abundances of the elements. A list of the selected windows for each element analyzed by the ASPCAP pipeline can be found in the DR16 reference paper by J\"onsson et al. (2020). We searched for updates in the $gf$-values for all atomic transitions in all windows of a given element. It should be kept in mind, however, that in many instances, due to blends with other molecular features, only some pixels spanning an atomic line carry the information that is fitted to derive the abundance of that particular species.  The first step here was to compile a list with the transitions that were present in the spectral windows used by ASPCAP in the determination of the different elemental abundances.  
With such a list in hand, we searched the National Institute of Standards and Technology Database (NIST; https://nist.gov/pml/atomic-spectroscopy-databases; Kramida et al. 2017), the Vienna Atomic Line Database (VALD; http://vald.astro.uu.se; Piskunov et al. (1995; VALD1) and Ryabchikova et al. (2015; VALD3), as well as the literature in general for updated sources for the $f$-values of the transitions for each individual element. The preferred sources for the $f$-values were experimental laboratory measurements, but when these were not available we used theoretical $f$-values with well-constrained uncertainties, or, if nothing else, the $f$-values from the Kurucz (2017) line list were used.

Table \ref{lab data} contains the considered species and the adopted sources of the $f$-value data for the atomic lines that ultimately served as input for the construction of the APOGEE line list. Whenever an $f$-value was taken from the NIST database we note it in the table (along with the original reference) and in such cases we adopt the NIST grading system for assessing the uncertainties in the $f$-values. The NIST grading system along with the corresponding percent uncertainties and associated errors in the $\log{(gf)}$ values are presented in Table \ref{uncertainties}. If the $f$-value was taken from a literature reference we used the uncertainties quoted in the studies themselves. 

The NIST database contained $f$-values for a number of the element transitions in the APOGEE spectral windows; in most cases these were laboratory measurements, although other references, such as TOPBASE (Opacity Project), and theoretical data with well constrained uncertainties are also included in the NIST compilation and used here.
The NIST $\log{(gf)}$ values for the H I, He I, C II, C III, C IV, N I, N II, N III, N V, O I, O II, O III, Mg II, Si II, and Si III transitions (with no windows in ASPCAP)  remained the same as in the previous APOGEE line list (we refer to Table 1 in Shetrone et al. 2015 for the specific references used).
For the two Na I transitions ($\lambda_{Air}$= 16373.853\AA\  and 16388.858\AA) in the two sodium windows, the NIST database provides accurate (accuracy=`A') $\log{(gf)}$ values from the NIST ``Multiconfiguration Hartree-Fock and Multiconfiguration Dirac-Hartree-Fock'' Database (non-orthogonal B-spline Configuration interaction calculations; Froese Fischer, downloaded in 2002).
For most of the Mg I transitions, we used the experimental $f$-values in Pehlivan Rhodin et al. (2017) or the Opacity Project $f$-values in Butler et al. (1993), while for three of the Mg I lines we relied on the Kurucz (2017) line list. For Al I lines the $f$-values are from the Opacity Project (Mendoza et al. 1995), although these have different levels of accuracy in the NIST Database.
For Si I, most of the $f$-values are from laboratory measurements by the Lund-Malm\"o atomic physics group (Pehlivan Rhodin 2018); for a few Si I lines unavailable in Pehlivan Rhodin (2018), we used the TOPBASE $gf$-values (Nahar \& Pradhan 1993), or relied on the Kurucz (2017) line list.
For the P I lines, we adopted the $gf$-values from the theoretical work by Bi\'emont et al. (1994), along with an estimated uncertainty in the $\log{(gf)}$-values of 0.1 dex. 
For the S I, K I and Ca I transitions we used the theoretical studies of Zatsarinny \& Bartschat (2006), Safranova et al. (2013), and Hansen et al. (1999), respectively.

Laboratory measurements of oscillator strengths are not available in the APOGEE spectral window for V I, Cr I, Co I, Ni I, and Cu I and we relied on the $gf$-values from the baseline Kurucz (2017) line list. 
There are many transitions of Fe I in the APOGEE windows and similarly there are no experimental data available for most of them, except for the several Fe I lines that were specifically studied to cover the APOGEE region in Ruffoni et al. (2013), and a few Fe I lines that were in Fuhr \& Wiese (2006). 
For manganese, we adopted the laboratory measurements from Blackwell-Whitehead et al. (2005), when available. For most of the Ti I lines we used the laboratory measurements of Blackwell-Whitehead et al. (2006) and Lawler et al. (2013); for one Ti I transition we used the Kurucz $f$-value.  
 Wood et al. (2014) provided a laboratory $f$-value measurement to support of the APOGEE Survey observations of the only useful line of Ti II in the APOGEE region ($\lambda_{Air}$=15873.84\AA).

\begin{deluxetable}{cl}
\tablecaption{
Sources for Oscillator Strengths \label{lab data}}
\tablehead{
\colhead{Species} & \colhead{Source} 
}
\startdata
C I &  NIST - Hibbert et al. (1993) \\
Na I & NIST - Froese Fischer (2002) \\
Mg I &  Pehlivan Rhodin et al. (2017); NIST - Butler et al. (1993); Kurucz (2017) \\
Al I &  NIST - Mendoza et al (1995) \\
Si I &  Pehlivan Rhodin (2018); NIST - Nahar \& Pradhan (1993); Kurucz (2017) \\
P I  &  Bi\'emont et al. (1994)   \\
S I & NIST - Zatsarinny \&  Bartschat (2006); Kurucz (2017) \\
K I & NIST - Safranova et al. (2013) \\  
Ca I & Hansen et al. (1999) \\
Ti I & NIST - Blackwell-Whitehead et al. (2006); Lawler et al. (2013); Kurucz (2017)  \\
Ti II &  Wood et al. (2014) \\
V I &  Kurucz (2017) 
\\
Cr I &  Kurucz (2017) \\
Mn I & Blackwell-Whitehead et al. (2005); Kurucz (2017) \\
Fe I & NIST - Fuhr \& Wiese (2006); Kurucz (2017) \\ 
Co I & Kurucz (2017) \\
Ni I & Kurucz (2017) \\
Cu I & Kurucz (2017) \\
Ce II & Cunha et al. (2017) \\
Nd II & Hasselquist et al. (2016) \\
\enddata    
\end{deluxetable}

\begin{deluxetable}{lll}
\tablecaption{
Oscillator Strength Uncertainties in NIST \label{uncertainties}
}
\tablehead{
\colhead{Grade} & \colhead{Uncertainty} & \colhead{$\delta\log{(gf)}$}
}
\startdata
AAA  & $\le	0.3\% $ &  $\pm$ 0.0013 \\
AA	 & $\le	1\% $   & $\pm$ 0.0043 \\
A+	 & $\le	2\% $  & $\pm$ 0.0086 \\
A	 & $\le	3\% $  & $\pm$ 0.0128 \\
B+	 & $\le	7\% $  & $\pm$ 0.0294 \\
B	 & $\le	10\% $ & $\pm$ 0.0414 \\
C+	 & $\le	18\% $ & $\pm$ 0.0719 \\
C	 & $\le	25\% $ & $\pm$ 0.0969 \\
D+	 & $\le	40\% $ & $\pm$ 0.146  \\
D	 & $\le	50\% $ & $\pm$ 0.176 \\
E	 & $>	50\% $ &  $\pm$ 0.300 \\
\enddata
\end{deluxetable}

\subsubsection{Heavy Element Lines in DR16}

One modification to the DR16 line list was the addition of a modest number of heavy-element ($Z\ge57$) lines that were identified in a combination of the spectra of Arcturus and/or the Sun, the FTS spectra of the ``standard'' cool giants, as well as s-process enriched red giants observed in APOGEE.  Three heavy elements were found to be detectable in the APOGEE window, with identified lines from Ce II (8 lines), Nd II (10 lines), and Yb II (1 line).  The initial identifications of these lines were facilitated by the compilation of heavy-element spectral lines in the wavelength range of $\lambda$10,000 $-$ 40,000\AA\ by Outred (1978).  By and large, these lines are mostly weak and usually blended in $R\sim22,500$ stellar spectra (although a few of the Ce II lines can be fairly strong in red giants); however, they can provide abundances in certain populations observed by APOGEE.  Cerium and neodymium are primarily products of neutron captures via the s-process, while ytterbium can be used to probe the r-process (about 68\% of solar system Yb is produced through r-process nucleosynthesis and 32\% from the s-process; Burris et al. 2000).

The oscillator strengths for the Ce II and Nd II lines were taken from the APOGEE studies by Cunha et al. (2017) and Hasselquist et al. (2016), respectively, and details of the line identifications and subsequent determinations of $gf$-values are described in detail in these references. Here we used the the $f$-value uncertainties quoted in those studies to constrain the astrophysical fitting.

\begin{figure}
\epsscale{0.60}
\includegraphics[width=0.70\textwidth,angle=0]{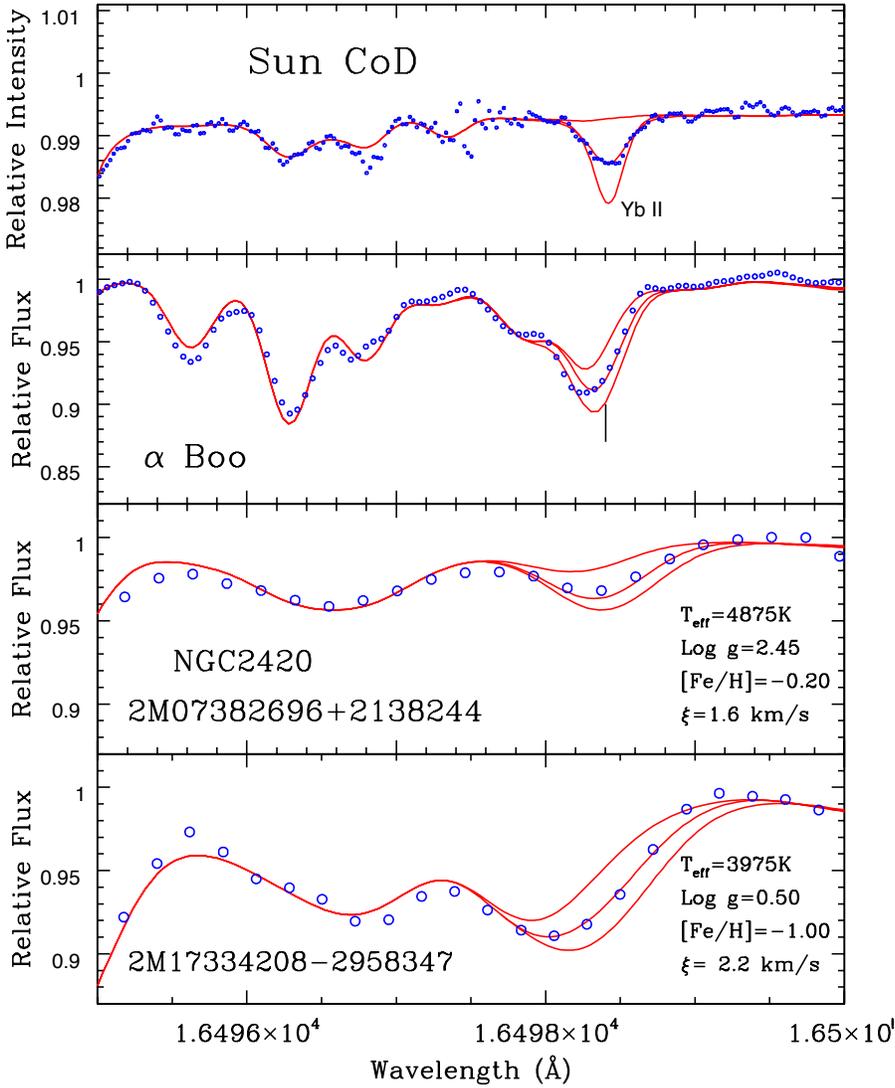}
\caption{An illustration of the Yb II line ($\lambda$16498.42\AA) and its behavior in four different types of stars: the Sun, the mildly metal-poor K-giant Arcturus, a clump giant in the old open cluster NGC2420, and a moderately metal-poor, cool, field red giant. The solar spectrum is Center-of-Disk (COD) from Livingston \& Wallace (1991), the Arcturus spectrum is the FTS atlas from Hinkle et al. (1995), while the bottom two red giant spectra are from APOGEE.  Each panel contains three synthetic spectra, with one having no Yb, another having the Yb abundance that fits the Yb II line, while a third has a Yb abundance larger by +0.2 dex compared to the best-fit abundance.  
\label{f2}}
\end{figure}

While the identifications and derivation of $gf$-values for the Ce II and Nd II lines were presented in Cunha et al. (2017) and Hasselquist et al. (2016), the details for Yb II are presented here.  The Yb II line identified in the APOGEE spectral window is at $\lambda_{Air}$=16498.42\AA\ with an excitation potential of $\chi$=3.017 eV and was found to be weak, but isolated in the Center-of-Disk (COD)
solar spectrum from Livingston \& Wallace (1991; top panel of Figure 2). Ytterbium in the solar system consists of seven stable isotopes, with 71\% consisting of five even isotopes all having nuclear spins of $I=0$, and two odd isotopes (29\%) with $I$ values of 5/2 and 1/2; however, since this Yb II line will always be quite weak, hyperfine and isotopic splitting are not included in modeling this line.

The solar system abundance of Yb remains somewhat uncertain, as Asplund (2009) lists a photospheric abundance of $A$(Yb) $=1.08$, while the corresponding meteoritic abundance is $A$(Yb) $=0.91$.  The Database on Rare Earths At Mons University (DREAM) lists a value of $\log{(gf)}=-0.64$ (http://hosting.umons.ac.be/html/agif/databases/dream.html; Bi\'emont et al. 1999), which was derived from lifetime calculations by Bi\'emont et al. (1998). 
Uncertainties in the lifetime calculations, as discussed by Bi\'emont
et al.  (1998), suggest an uncertainty for this gf-value of $\sim\pm$0.06
dex.  This gf-value yields a solar abundance (see the top panel of Figure
2) of $A$(Yb) $=1.03$, that falls nearly midway between photospheric and
meteoritic values, so the value of $\log{(gf)}=-0.64$ is used in the base
atomic line list.  The Yb II line is quite clean in the Sun, but weak, as
shown in the top panel of Figure 2.  At the resolution of the APOGEE spectra
($\sim$0.7\AA), test synthetic spectra show that the line-depth becomes too
small to be of practical use in measuring Yb abundances in solar-type dwarf
stars observed as part of APOGEE.

The overall behavior of the Yb II line in different types of red giants is shown in the bottom three panels of Figure 2.
As an illustration of the sensitivity of the Yb II line to the ytterbium abundance, each panel in Figure 2 includes three synthetic spectra having differing Yb abundances: one with no Yb, a second with a best-fit Yb abundance, and a third that has a Yb abundance larger than the best-fit abundance by +0.2 dex.  Arcturus, shown in the second panel, is an example of a mildly metal-poor K2 giant where the Yb II line is buried within the (7-4)R59 $^{12}$C$^{16}$O line.  In the case of a well-studied standard star, such as Arcturus, where the carbon and oxygen abundances are well-determined, the (7-4)R59 feature appears too deep and the addition of a Yb abundance with [Yb/Fe]=+0.07 reproduces the feature well.  The internal uncertainty in the [Yb/Fe] ratio is set by uncertainties in the Yb II gf-value ($\pm$0.06 dex) and the uncertainty in the Arcturus Fe-abundance ($\sim\pm$0.05 dex), or $\pm$0.08 dex if the respective uncertainties are added in quadrature.  In similar red giants within the APOGEE survey, observed at lower resolution, the determinations of Yb abundances from this single line will be somewhat more uncertain; attempts to determine Yb abundances from APOGEE are favored in warmer red giants (where CO absorption is weaker) and metal-poor, low-gravity giants (where both CO formation is weakened by metallicity-squared and ionized lines are stronger).  The third panel in Figure 2 illustrates a red clump (RC) giant member of the open cluster NGC2420 (RC giants are common APOGEE targets), 2M07382696+2138244, in the Yb II region, with the Yb II line yielding an abundance of [Yb/Fe]=+0.05.  In RC giants, the CO contamination is less than in the cooler giants, with a typical uncertainty in the Fe-abundance of about 0.1 dex, resulting in an estimated internal uncertainty in [Yb/Fe]$\sim\pm$0.12 dex.  The bottom panel shows the Yb II line in 2M17334208-2958347, a cool moderately metal-poor giant with a Yb II line that is fit for an abundance of [Yb/Fe]=+0.15$\pm$0.12. 

Another heavy element abundance that appears in DR16 results is for rubidium ($Z=37$), which is produced as a result of both the r- and s-processes, with the solar system mixture being 50\%/50\% (Burris et al. 2000).  There is a Rb I doublet in the APOGEE window at air wavelengths of $\lambda$15288.430\AA\ and $\lambda$15289.480\AA\ having $\chi=1.579$eV (Sansonetti 2008).  The wavelengths in the Kurucz line list for these transitions are somewhat different, with air wavelengths of $\lambda$15288.938\AA\ and $\lambda$15289.966\AA, respectively.  Tests of the presence of these lines in Arcturus and in the M-giants $\beta$ And and $\delta$ Oph indicate that the Sansonetti (2008) wavelengths are preferred, although only the stronger Rb I line at $\lambda$15289.480\AA\ is detected.  This line is weak, but may be detectable in APOGEE spectra of s-process rich populations, or in the cooler red giants.  In addition, there are non-LTE departure coefficients calculated for these Rb I transitions by Korotin (2020).

\subsection{Molecular Lines}

Molecules in the DR16 list include CO, OH, CN, C$_{2}$, H$_{2}$O, FeH, and SiH, with the data source for each molecule discussed below and listed in Table \ref{molecular}.

\begin{deluxetable}{cl}
\tablecaption{
Molecular Line lists \label{molecular}}
\tablehead{
\colhead{Molecule} & \colhead{Source} 
}
\startdata
CO & Li et al. (2015) \\
OH &  Brooke et al. (2015) \\
CN &  Sneden et al. (2014) \\
H$_{2}$O & Barber et al. (2006)  \\
C$_{2}$ & Yurchenko et al. (2018b) \\
FeH & Hargreaves et al. (2010)   \\
SiH & Yurchenko et al. (2018a) \\
\enddata    
\end{deluxetable}

\subsubsection{CO}

Rovibrational lines of CO are the primary carbon-abundance indicators in the APOGEE red giants (with C/O $\le1$) and cool dwarfs, with a number of vibration-rotation band systems running across the APOGEE spectral window, the strongest of which have $\Delta\nu$ = 3.  The previous DR14 line list used CO data from Goorvitch (1994), while in DR16 the CO data have been updated using wavelengths, excitation potentials, and $gf$-values from Li et al. (2015). Due to the importance of the CO lines in the derivation of stellar atmospheric parameters, as well as derived carbon abundances, it is useful to compare the CO data from Li et al. (2015) with that of Goorvitch (1994). 
The energy levels compare very well, with differences less than 0.001 eV, while the wavelength differences are typically a few m\AA.  There are small, systematic differences in the $gf$-values, and these are illustrated in Figure 3 as $\Delta\log{(gf)}$(Li et al. $-$ Goorvitch) as a function of excitation potential, $\chi$.  Only the detectable CO lines are shown, with $\Delta\nu=3$; there are higher-level CO lines (with $\Delta\nu\ge4$) that can have larger offsets between Li et al. and Goorvitch (of $\sim$+0.3 dex), but these lines have such large excitation potentials that they are too weak to affect the spectra of red giants and cool dwarfs.  In the stronger detectable CO lines that are the major C-abundance indicators, the Li et al. (2015) $gf$-values are slightly smaller than those from Goorvitch (1994), with a mean difference and standard deviation of $\Delta\log{(gf)}$(Li et al. - Goorvitch)$= -0.06\pm0.03$ dex.

Besides the main CO isotopic combination of $^{12}$C$^{16}$O, additional minor isotopic combinations are also in the DR16 line list, including $^{13}$C$^{16}$O, $^{12}$C$^{17}$O, and $^{12}$C$^{18}$O.

\begin{figure}
\epsscale{0.90}
\plotone{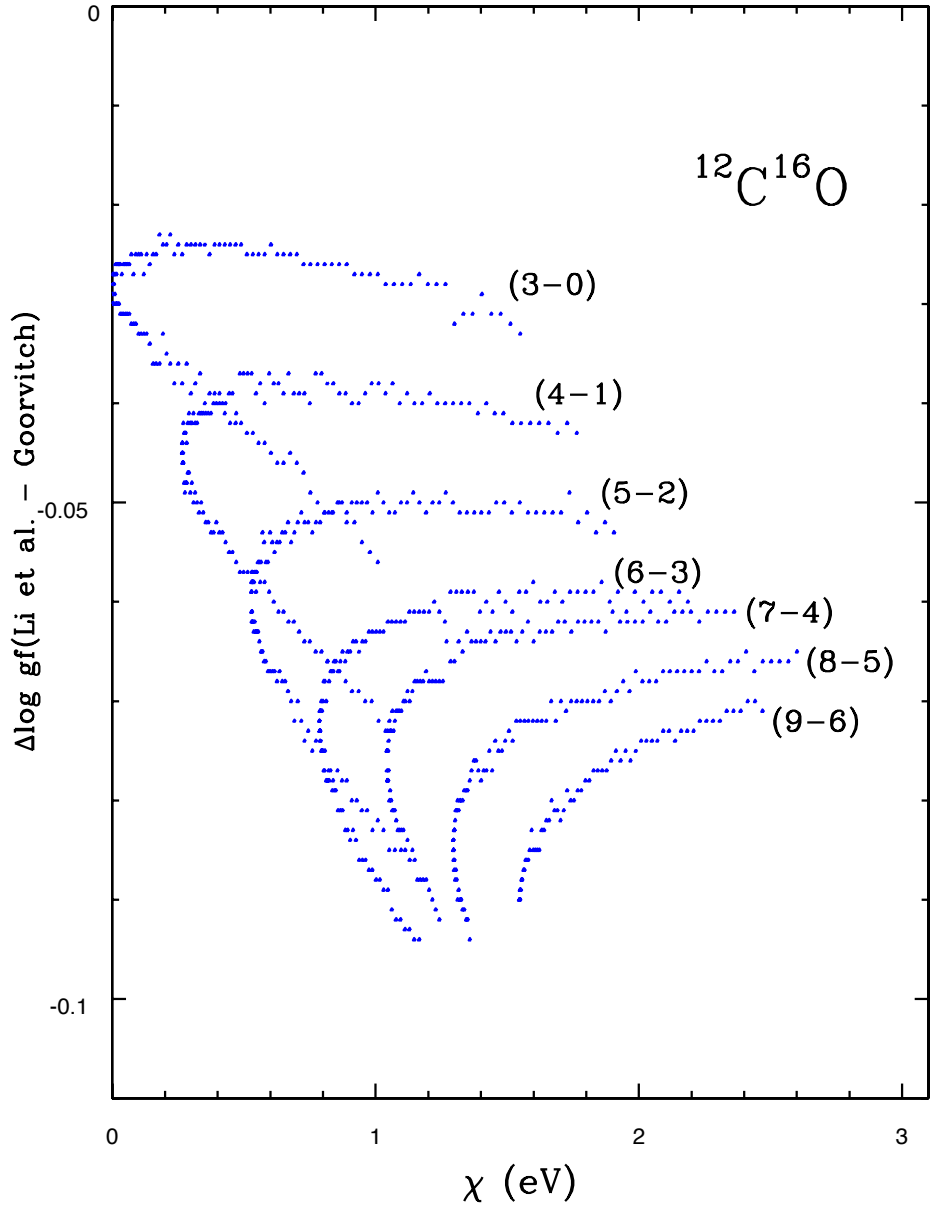}
\caption{Comparisons of $^{12}$C$^{16}$O $gf$-values between Li et al. (2015) and Goorvitch (1994), shown as $\Delta\log{(gf)}$ (Li et al. - Goorvitch) as a function of excitation potential ($\chi$).  The progression of vibration-rotation bands, from (3-0) to (9-6), which run across the 1.5 - 1.7$\mu$m, are evident, with the differences becoming greater for the higher vibration-rotation bands.
\label{f3}}
\end{figure}

\subsubsection{OH}

As with CO, a number of rovibrational lines from first-overtone bands of OH are found within the APOGEE window and abundances from the critical element oxygen rely on these OH lines.  In both the cool red giants and the M-dwarfs, the OH lines are some of the stronger features, remaining detectable down to very low metallicities, and are thus useful spectral lines in metal-poor populations.

Spectral-line data for OH were updated for DR16 using Brooke et al. (2016); the DR14 line list used data from Goldman et al. (1998) for OH.  A comparison of the $gf$-values between Brooke et al. (2016) and Goldman et al. (1998) is shown in Figure 4, as $\Delta$(log gf(Brooke et al. $-$ Goldman et al.)) versus excitation potential, and a systematic shift between the two OH studies can be seen; the Brooke et al. values are slightly smaller, on average, when compared to Goldman et al. (1998), with a mean difference, for the majority of detectable lines, of $\Delta$(log(gf)= -0.08 dex.  This difference becomes greater for the higher $J$-values and $\Delta\nu=3$ second-overtone bands, however these lines become progressively weaker and do not impact significantly derived O-abundances.
The minor oxygen isotopes are also included in the OH list ($^{17}$OH and $^{18}$OH).

\begin{figure}
\epsscale{0.90}
\plotone{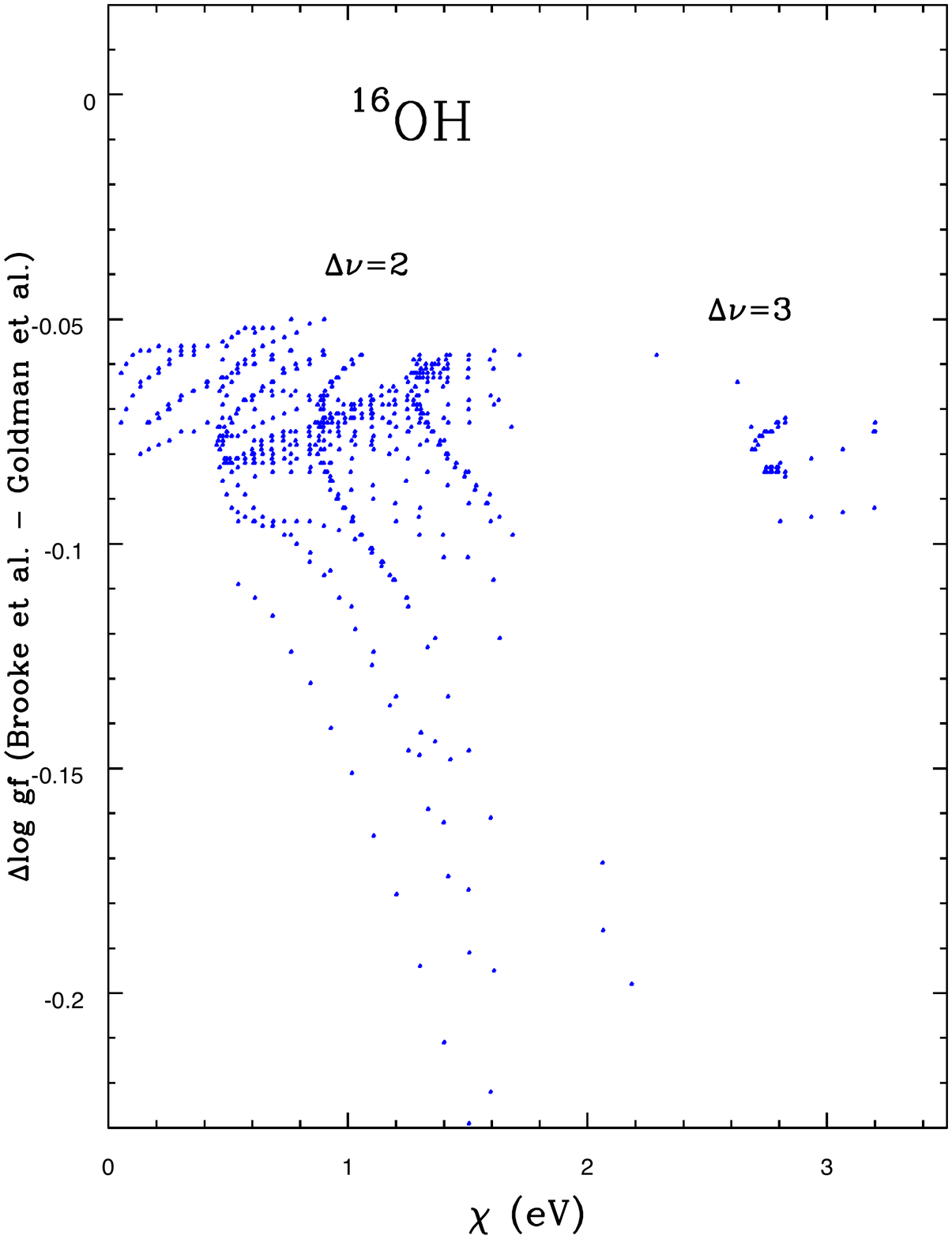}
\caption{Comparison of Brooke et al. (2015) and Goldman et al. (1998) $^{16}$OH $gf$-values for lines falling in the APOGEE window.  Absorption from OH is dominated by $\Delta\nu$=2 transitions, with the structure in the $\log{(gf)}$ comparisons arising from the (in order of increasing $\chi$) (3-1), (4-2), and (5-3) vibration-rotation systems (the lines at very low $\chi$ are (2-0) lines). 
\label{f4}}
\end{figure}

\subsubsection{CN}

There was no change in the CN spectral-line source between DR14 and DR16, with the CN data taken from Sneden et al. (2014).  In combination with CO and OH, the CN lines close the constraints on the individual C, O, and N abundances, with these three diatomic molecules providing most of the CNO abundance information.

The minor isotopic lines of $^{13}$C$^{14}$N are included in the line list.

\subsubsection{C$_{2}$}

The recent study by Yurchenko et al. (2018b) was the source for the C$_{2}$ data, including wavelengths, excitation energies, and $gf$-values. The C$_{2}$ lines are, in general, weak in the oxygen-rich red giants, which are the primary targets in APOGEE; however, there are significant numbers of carbon stars in which the C$_{2}$ lines become quite strong and serve as carbon-abundance indicators.  The minor isotope $^{13}$C was included as both $^{13}$C$^{12}$C and $^{13}$C$^{13}$C.

\subsubsection{H$_{2}$O}

In preparation for DR16 and the inclusion by APOGEE of cooler red giants ($T_{\rm eff} \le 3500$K) as well as increasing numbers of M-dwarfs, many of which are exoplanet-hosts, H$_{2}$O lines were added to the DR16 line list (which were missing from DR14); absorption by numerous water lines becomes increasingly important in the cooler red giants, as well as in all M-dwarfs.

Data for the H$_{2}$O energy levels and Einstein A-values (from which $gf$-values were computed) for the various types of transitions were taken from Barber et al. (2006) and were used to construct the H$_{2}$O line list between $\lambda$15,000 to 17,000\AA; within this wavelength interval, the Barber data result in the generation of over 26 million water lines (using only one isotopic combination of $^{1}$H$_{2}^{16}$O, with deuterated water and the minor isotopes of $^{17}$O and $^{18}$O considered to be insignificant sources of H$_{2}$O absorption).

Including this large number of H$_{2}$O lines in the spectral synthesis calculations needed to generate the large number of library spectra is impractical, so the H$_{2}$O lines were culled to remove the weakest lines.  A very large fraction of the water lines are extremely weak and a so-called ``Boltzmann cut'' was applied to remove the weakest lines; the relative strength of a spectral line from a particular species depends on the combination of $\log{(gf\lambda)}$, temperature, and excitation potential, $\chi$, with the relative line strength, $S$, often written as $S= \log{(gf\lambda)} - \theta\chi$, where $\theta=5040/T$, with $\chi$ having units of eV and T in Kelvin, and here $\lambda$ is normalized to 16,000\AA.
Using an excitation temperature of $T=3500$K, two H$_{2}$O line lists were prepared, with one containing only transitions with $S>-8.5$ and another with only $S>-9.5$, with these stronger H$_{2}$O lines used in calculating the spectral libraries (these were the Boltzmann cuts).  These particular cuts were used after tests with synthetic cool-dwarf spectra, in which spectra computed with and without the cut were compared and found to differ by less than 1\% in flux.  The criteria for the inclusion of either the $S>-8.5$ list, or the $S>-9.5$ list, or no inclusion of water lines at all were as follows: (1) if T$_{\rm eff}>$4000K, H$_{2}$O lines were not included in the library spectra, (2) if T$_{\rm eff}$=3250K - 4000K, or [M/H]+[$\alpha$/M]$>$-1.5, the S$>$-8.5 line list was used, and (3) if T$_{\rm eff}<$3250K, the S$>$-9.5 line list was used.  In the end, these culled lists of H$_{2}$O lines added to the DR16 line list were found to fit and match very well the H$_{2}$O absorption in the M-dwarfs analyzed in Souto et al. (2017; 2018; 2020).

In addition to synthesizing and matching H$_{2}$O absorption features in the M-dwarfs as a test of the DR16 line list, transitions generated by the Barber et al. (2006) energy levels can be compared with data for H$_{2}$O lines from the more recent study by Polyansky et al. (2018), which appeared after the freezing of the DR16 line list.  Polyansky et al. (2018) included higher-energy levels, up to a maximum energy of 41,000 cm$^{-1}$ (5.08 eV), while Barber et al. (2006) included energy levels up to a level energy of 30,000 cm$^{-1}$ (3.72eV).  
The inclusion of higher energies in Polyansky et al. (2018) led to a significant increase in the number of H$_{2}$O energy levels (810,269) compared to Barber et al.'s (2006) total of 221,097 levels. 
This leads to many more transitions in Polyansky et al. (2018), although the higher-energy transitions are typically weak, or undetectable in the APOGEE targets.  Polyansky et al. (2018), in their Figure 2, compare their line list with that of Barber et al. (2006) as a plot of H$_{2}$O absorption, calculated for $T_{exc}=4000$K, versus wavenumber.  This comparison reveals that, in the infrared, the two line lists produce absorptions that are quite similar, including over the APOGEE wavenumber regime ($\sigma$=5880-6667 cm$^{-1}$), while at higher wavenumbers, particularly in the optical ($\sigma>$15,000 cm$^{-1}$), the absorption differences become much larger.

Differences between the Barber et al. (2006) and Polyansky et al. (2018) H$_{2}$O line lists across the APOGEE spectral window were investigated using cool stellar atmosphere models and comparing the synthetic spectra, an example of which is shown in Figure 5.  This particular example is for a model atmosphere having $T_{\rm eff}=3500$K, $\log{g}=4.5$, [Fe/H] $=0.0$, and $\xi=1.0$ km-s$^{-1}$, while the wavelength interval chosen to plot is near where the H$_{2}$O absorption is relatively strong (note, only H$_{2}$O lines are included in the synthesis in order to facilitate a direct comparison).  Differences between the two line lists are not large, being typically less than a few percent.  Quantitative comparisons to observed M-dwarf APOGEE spectra indicate that the more recent Polyansky et al. (2018) data provide a marginally better fit than the older Barber et al. (2006) data, although the fits are not significantly better.  Testing of the H$_{2}$O line lists will continue towards the final planned data release (DR17).

\begin{figure}
\epsscale{0.90}
\plotone{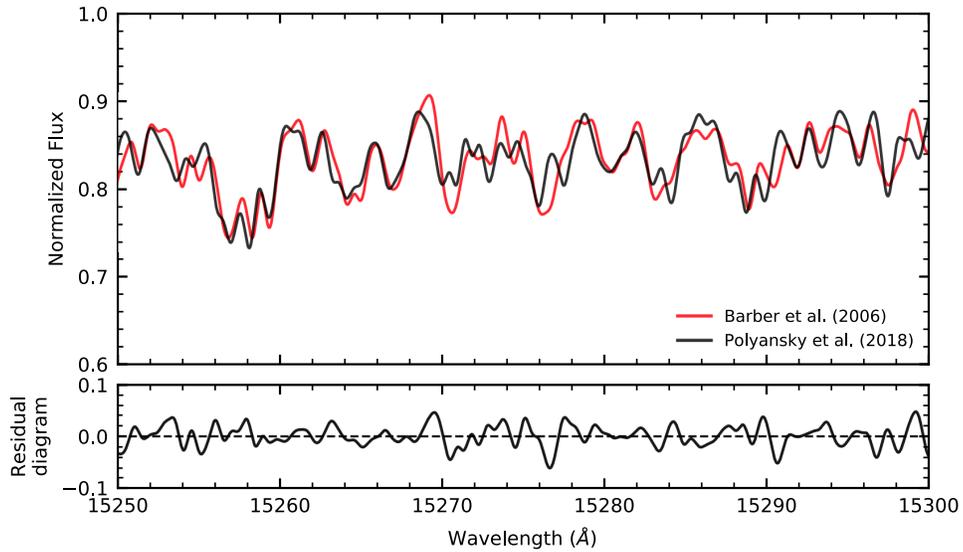}
\caption{A comparison of an H$_{2}$O synthetic spectrum generated with the DR16 Barber et al. (2006) line list (top panel, red line) with one computed from the Polyansky et al. (2018) list (black line).  Differences, in the sense of Barber$-$Polyansky, are shown in the bottom panel.  There are rather small ($\sim$0-5\%) differences in absorption, as well as slight differences in wavelength, which are not due to differences in $gf$-values for the same transitions (which agree), but due to the additional lines from higher excitation energies that are included in the Polyansky et al. (2018) list. The synthetic spectra were computed with a model atmosphere having $T_{\rm eff}=3500$K, $\log{g}=4.5$, and [Fe/H] $=0.00$. 
\label{f5}}
\end{figure}

\subsubsection{FeH}

The DR14 APOGEE line list did not include transitions from iron hydride, although FeH is an important contributor to the H-band APOGEE spectra of M-dwarfs (Souto et al. 2017).  Spectral lines from FeH were added to DR16 using data from Hargreaves et al. (2010), which consist of lines arising from E$^{4}\Pi$-A$^{4}\Pi$ electronic transitions from $\lambda$1.58$\micron$ up to $\lambda$1.7$\micron$.  Wallace \& Hinkle (2001) identified lines from this transition in the spectra of sunspots, along with M and L dwarfs, and these FeH lines match closely the wavelengths of lines in M-dwarf APOGEE spectra that were missing from the DR14 syntheses. 

Hargreaves et al. (2010) do not provide $gf$-values for the FeH lines, so Souto et al. (2017) computed $gf$-values using the line intensities provided by Hargreaves et al. (2010) and the expression for converting HITRAN-like intensities to Einstein $A$-values from Simeckova et al. (2006; see their Equation 20).  The $A$-values were then converted to $gf$-values using the standard expression from Larsson (1983):
$$gf = (1.499 (2J+1) A)/(\sigma^{2})$$
\noindent where $A$ is the Einstein $A$-value (in s$^{-1}$), $J$ is the lower-state angular momentum, and $\sigma$ is the wavenumber (in cm$^{-1}$).  The gf-values presented in Souto et al. (2017) were adopted for the DR16 line list.

\subsubsection{SiH}

Although SiH lines are expected to be weak (and undetectable) in virtually all of the APOGEE red giants and dwarfs, these molecular lines were included in the line list.  The data were taken from Yurchenko et al. (2018) and include the isotopes of $^{28}$SiH, $^{29}$SiH, and $^{30}$SiH.

\subsubsection{Molecular Absorption Comparisons in APOGEE Spectra of a Typical Red Giant and M-Dwarf} 

\begin{figure}
\epsscale{1.00}
\rotate{270}
\includegraphics[width=0.80\textwidth,clip,angle=0]{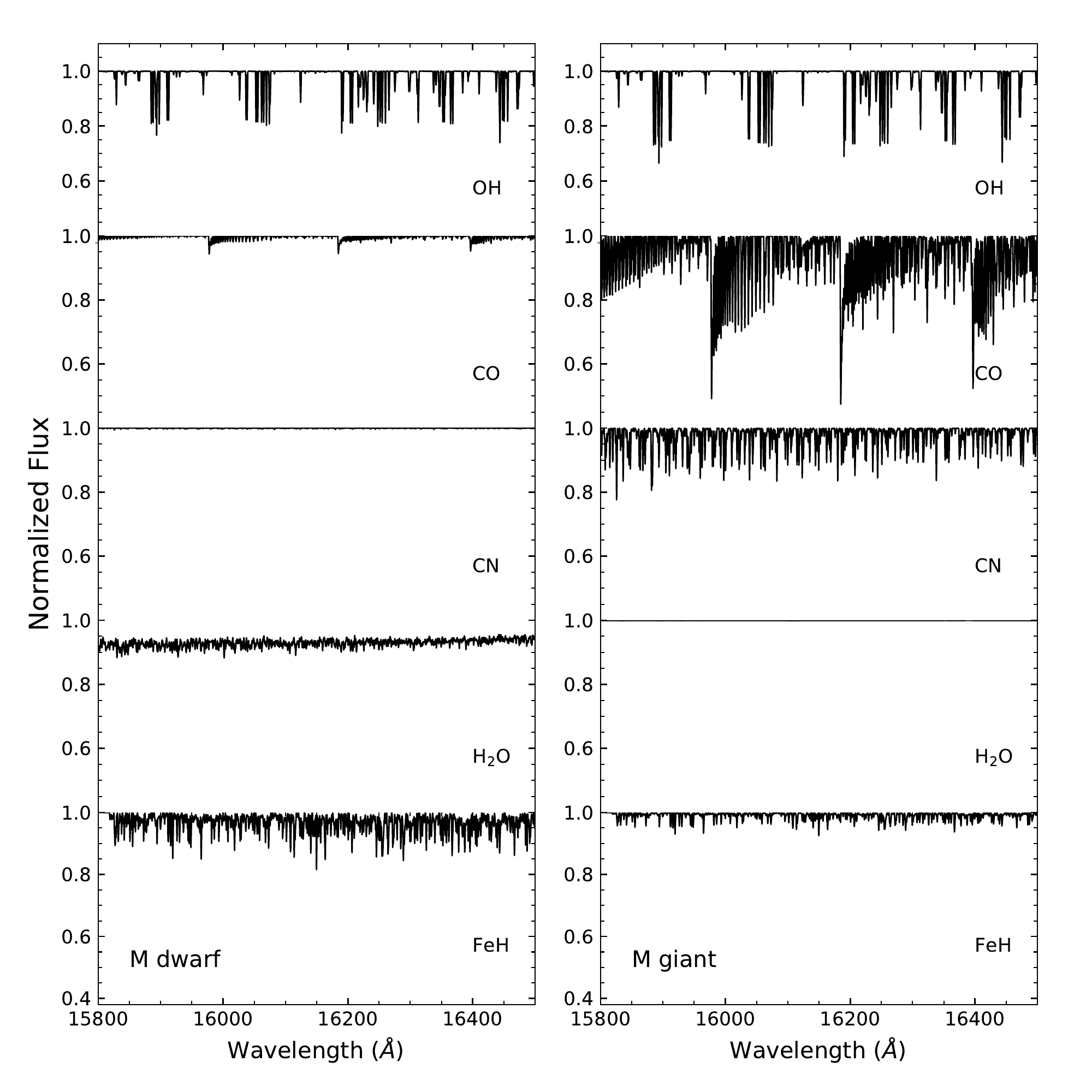}
\caption{An illustration of absorption from the major molecular species (CO, OH, CN, H$_{2}$O, and FeH) in typical red giant ($T_{\rm eff}=3800$K, $\log{g}=1.2$, $\xi=2.0$ km-s$^{-1}$, [Fe/H] $=0.0$) and M-dwarf ($T_{\rm eff}=3500$K, $\log{g}=4.4$, $\xi=1.0$ km-s$^{-1}$, [Fe/H]$ =0.0$) spectra. 
\label{fig:f6}}
\end{figure}

As an aid to visualize the relative importance of the various molecules in shaping the absorption-line spectra in cool giants and dwarfs, molecular synthetic spectra are illustrated in Figure \ref{fig:f6} for the more significant molecules (OH, CO, CN, H$_{2}$O, and FeH), with an isolated spectrum for each molecule.  The spectra were computed for models of a typical APOGEE  M-giant and M-dwarf.  This figure provides a sense of how much each molecular species influences the red giant and red dwarf spectra, respectively.

\section{Astrophysical Adjustment to the Line Parameters}

As shown previously by Shetrone et al. (2015), the combined list of theoretical and laboratory $gf$-values and damping constants for the atomic spectral lines that are strong enough to be clearly detected and well-sampled can be improved by comparisons with high-resolution, high-quality spectra of well-studied standard stars, i.e., by applying ``astrophysical'' adjustments.  Note that adjustments are limited to atomic lines only and no adjustments were made to the molecular lines discussed in Sections 2.2.1 to 2.2.7.

\subsection{The Astrophysical Adjustment Methodology}

We follow the approach used by Shetrone et al. (2015): spectral line parameters are adjusted for well-defined lines in the model spectra of the Sun and the K-giant Arcturus ($\alpha$ Boo) in order to match the observed lines in high-resolution, high-$S/N$ FTS spectra within the limits of the laboratory (or theoretical) parameter uncertainties.  

The same line-matching code as used by \citet{shetrone15} and described in \citet{bizyaev15} was adopted, as were the same reference FTS spectra used in previous APOGEE line lists: the center-of-disk (COD) solar spectrum from \citet{livingston91}, and the NIR spectral atlas of Arcturus by \citet{hinkle95}.
A major change with respect to \citet{shetrone15} is that the synthetic spectra were generated with TurboSpectrum \citep{TS,TS15}, a local thermodynamic equilibrium (LTE), spectral synthesis code that can be used in both plane-parallel and spherical geometries. 
Turbospectrum version 15.2 (the last TurboSpectrum release at the time of these calculations) was used.  TurboSpectrum was also the code used to compute the synthetic spectral libraries for DR16 (J\"onsson et al. 2020).

The LTE MARCS code (Gustafsson et al. 2008) was used to construct the stellar atmosphere models spanning 56 depth points. For consistency, standard MARCS physical parameters were adopted: 1.5 for the mixing length parameter, 0.076 for the temperature distribution within the convective elements, and 8 for the energy dissipation by turbulent viscosity. 
The model atmosphere for the Sun was computed in plane-parallel geometry, while the model for Arcturus was computed in spherical geometry assuming a mass of 0.7M$_\odot$.
The models were also computed with self-consistent chemical abundance distributions.  

We use a microturbulent velocity of 0.7 \kms\ for the COD synthetic spectra of the Sun (e.g., Blackwell et al. 1995). The spectra were convolved with a Gaussian kernel that corresponds to 1.20 \kms, which helps account for the instrumental profile. 
We adopt a microturbulent velocity of 1.7 \kms\  for Arcturus (Ramirez \& Allende Prieto 2011). 
Synthetic spectra of Arcturus were convolved with a rotational profile for $v\sin{i} = 2$ \kms \citep{gray81,gray06}, calculated with the limb darkening parameter of 0.48 \citep{claret00}, and with a Gaussian kernel of $\sigma$=1.9 \kms\ to correct the spectra for both the rotational and resolution effects. 

For the Sun, the abundances are from Grevesse, Asplund \& Sauval (2007). The abundances adopted for Arcturus were from a small number of studies, with most of the abundances taken from Ramirez \& Allende Prieto (2011); the abundances adopted for the Sun and Arcturus are summarized in Table \ref{tab1}.  In addition to Ramirez \& Allende Prieto (2011), the CNO abundances for Arcturus were derived here as part of the DR16 line list upgrade, using the same OH, CO, and CN lines discussed Section 2. 

The CNO abundances derived here for Arcturus can be compared to two recent detailed analyses of CNO in Arcturus by Abia et al. (2012) and Sneden et al. (2014).  Abia et al. (2012) analyzed the same FTS spectrum of Arcturus used here, but focused their analysis on the wavelengths $\lambda$4.55 - 5.56$\mu$m, which also contains transitions of OH, CO, and CN.  Differences in the CNO abundances derived here compared to Abia et al. are small, with $\Delta = {\rm APOGEE} - {\rm Abia\ et\ al.\ (2012)}$ being $\Delta(A(^{12}$C)) $=-0.03$, $\Delta(A(^{14}$N)) $=-0.05$, and $\Delta(A(^{16}$O)) $=-0.13$.  In both studies the carbon isotope ratios are similar, with $^{12}$C/$^{13}$C $=7\pm$1 derived here and 9$\pm$2 from Abia et al. (2012).  A comparison to Sneden et al. (2014) is particularly interesting, as their analysis is based on the optical spectrum of Arcturus, with the carbon abundance determined using C$_{2}$ Swan bands at 4737\AA\ and 5135\AA, as well as the CH G-band around 4270-4330\AA, the nitrogen abundance from red CN lines in spectral intervals from $\lambda$6000-11,000\AA, and the oxygen abundance from the [O I] line at $\lambda$6300.27\AA.  The Arcturus CNO abundances derived here match those of Sneden et al. (2014) well, with $\Delta(A(^{12}$C)) $=+0.01$, $\Delta(A(^{14}$N)) $=-0.04$, and $\Delta(A(^{16}$O)) $=+0.01$.  Sneden et al. (2014) also derive $^{12}$C/$^{13}$C $=7\pm$1 from CN lines near $\lambda$8000\AA.  Taken together, these three independent determinations of CNO in Arcturus suggest that the abundances of these important elements are well-constrained in this standard red giant.

\begin{deluxetable} {rcccl}
\tablecaption{The Adopted Elemental Abundances for the Sun and Arcturus. \label{tab1} }
\tablehead{ \colhead{Z} & \colhead{Elem.} & \colhead{Sun} & \colhead{Arcturus} & \colhead{Reference} 
}
\startdata
6  & C  & 8.39 & 8.03 & Smith (2017) \\
7  & N  & 7.78 & 7.62 & Smith (2017) \\
8  & O  & 8.66 & 8.63 & Smith (2017) \\
11 & Na & 6.17 & 5.76 & Ramirez \& Allende Prieto (2011) \\
12 & Mg & 7.53 & 7.38 & Ramirez \& Allende Prieto (2011) \\
13 & Al & 6.37 & 6.19 & Ramirez \& Allende Prieto (2011) \\
14 & Si & 7.51 & 7.32 & Ramirez \& Allende Prieto (2011)  \\
15 & P  & 5.36 & 5.11 & Maas et al. (2017) \\
16 & S  & 7.14 & 6.94 & Ryde et al. (2009) \\
19 & K  & 5.08 & 4.76 & Ramirez \& Allende Prieto (2011) \\
20 & Ca & 6.31 & 5.90 & Ramirez \& Allende Prieto (2011) \\
21 & Sc & 3.05 & 2.84 & Ramirez \& Allende Prieto (2011) \\ 
22 & Ti & 4.90 & 4.65 & Ramirez \& Allende Prieto (2011) \\
23 & V  & 4.00 & 3.54 & Wood et al. (2018) \\
24 & Cr & 5.64 & 5.07 & Ramirez \& Allende Prieto (2011) \\
25 & Mn & 5.39 & 4.66 & Ramirez \& Allende Prieto (2011) \\
26 & Fe & 7.45 & 6.93 & Ramirez \& Allende Prieto (2011) \\
27 & Co & 4.92 & 4.49 & Ramirez \& Allende Prieto (2011) \\
28 & Ni & 6.23 & 5.77 & Ramirez \& Allende Prieto (2011) \\
29 & Cu & 4.21 & 3.71 & Ramirez \& Allende Prieto (2011) \\
58 & Ce & 1.58 & 0.99 & Cunha et al. (2017) \\
60 & Nd & 1.45 & 0.94 & Overbeek et al. (2016) \\
70 & Yb & 1.08 & 0.63 & This study \\
\enddata
\end{deluxetable}

The compiled line list described in \S2 and covering the wavelength range from $\lambda$15,000 to 17,000\AA\ was used as input for the line-matching code.  TurboSpectrum was used to generate spectra for individual atomic lines in the list, with the line depth with respect to the continuum being used to select spectral lines that are strong enough in either, or both, the spectra of Arcturus and the Sun to be used to define astrophysical adjustments. 
Given the different quality of the reference spectra, lines deeper than 0.003 in the Sun and 0.01 in Arcturus were included as potential targets for astrophysical adjustments if the  uncertainties in the $gf$-values were known.  We account for less reliability of lines without known uncertainties and, in this case, keep only the lines deeper than 0.0075 and 0.025 in the Sun and Arcturus, respectively.  The criteria described above result in a list of 1664 atomic lines measured for astrophysical adjustments in at least one of the standard stars. 
We also investigated 250 lines deeper than 0.15 of the continuum level in the solar spectrum for damping constant adjustments (the shallower lines do not have enough signal in their wings for this task). 

The lines in the list were adjusted in order of their depth, starting with the deepest ones. For each line, a 0.8\AA~ wavelength range was considered, unless the line had HFS components. In the latter case, the wavelength range was extended 0.4\AA\ beyond the bluest and reddest components of the HFS families.  Parameters for the entire HFS family were changed simultaneously during the fitting.  The line position was allowed to vary by up to 0.25 \kms\ (or up to 0.013\AA\ at 16,000\AA), while the $\log{(gf)}$ value could vary by up to twice its estimated uncertainty in the course of the least squares optimization, which used 
the downhill simplex algorithm \citep{amoeba}. The spectra of the Sun and Arcturus were considered separately for each line of interest, and the resulting $\log{(gf)}$ value is the weighted mean over both stars, where the weight is the line depth; this was the same method employed by \citet{shetrone15}.  The effects of strong line wings were taken into account by computing synthetic spectra over a wider range ($\pm$9\AA) before cutting out the smaller piece for fitting.   

Some of the spectral lines that were adjusted had no known or estimated log(gf) uncertainties and the astrophysical adjustments to these lines were constrained to within $-$2.0 and +0.75 dex from the original $\log{(gf)}$.  Iterations on the damping parameters for the strong solar lines were restricted to $\pm$0.4 dex changes, with a 3.0\AA~ wavelength range being considered around the lines of interest. 

The iterative process for the line parameter adjustments includes using the resulting line list as input for a next iteration. We run several iterations of the line parameter adjustments as follows: first, two $\log{(gf)}$ iterations were performed with the solar spectrum. Next, two damping constant iterations were done on the strong solar lines.  After that, six more $\log{(gf)}$ iterations, using both the Sun and Arcturus were completed. 
Although the evaluation of a single line does not need the iterative adjustment, it is required for lines in blends.  While most of the line parameter adjustments converge quickly, $\log{(gf)}$ values for some lines do not settle down even with a large number of iterations. This occurs when two strong lines are located near each other in wavelength. We found 36 such lines that continue to have large $\log{(gf)}$ adjustments after all iterations.  In these cases, the original $\log{(gf)}$ values were replaced with the average values over the second to sixth iterations, which was found to provide good final fits. 

The procedure used to derive the astrophysical adjustments resulted in significantly reduced difference between synthetic and observed spectra for both the Sun and Arcturus with this single iterated line list. Figure 7 shows these differences as a function of wavelength; each point represents the RMS value between the synthetic and observed spectra over 20\AA\ chunks, with a synthesis step of 0.02\AA.  The initial differences are shown as red and the final interated differences are in blue; in both the Sun and Arcturus, the initial differences are significantly larger than the final differences, with the single interated line list providing superior fits to both standard stars.  We note again that changes to the $gf$-values of the atomic lines are limited to values that are less than twice the uncertainties.  In the case of the Sun, the mean RMS value is 0.017 initially and is reduced to 0.008 in the final line list, while for Arcturus the corresponding mean RMS values are 0.026 and 0.018, respectively.  Overall fits are better for the Sun when compared to Arcturus, as the solar spectrum displays far fewer lines and much less spectral-line absorption than Arcturus (there are larger stretches of ``line-free continuum'' in the Sun).  Overall in the APOGEE window, Arcturus has about 2.2 times more absorption than the Sun (122\AA\ of equivalent-width absorption, in total, compared to 55\AA\ for the Sun).

Differences between the observed and the synthetic solar and Arcturus
spectra generated by the final iterated line list were used to identify
mismatched regions which were then not used in fitting library spectra to
observed spectra in ASPCAP. These mismatched regions were excluded
from the fits to library spectra by use of a global mask, consisting of a
weight of either 0.0 or 1.0 for each wavelength (pixel).  Wavelengths were
identified where the difference between the observed and synthesized spectra
of Arcturus differ by more than 0.05 in normalized flux, or where the
difference between the observed and synthesized spectra of the Sun differ by
more than 0.03 in normalized flux; the pixels at these wavelengths were
given zero weight in the fits, while the other pixels were assigned a weight
equal to 1.0. The comparison was made using spectra smoothed to APOGEE
resolution.  Some regions masked by this algorithm were removed because they
were considered to be residuals of telluric lines still present in the
observed spectra.  After visual inspection, one mask covering a region
$\lambda$16040-16050\AA\ was widened somewhat, and one covering the
Brackett-11 hydrogen line at $\lambda16806.5\pm15$ \AA~was added manually. 
This process resulted in 21 regions of the spectra being masked out in all
DR16 ASPCAP-fits of both parameters and abundances.  For completeness,
the global mask wavelengths and weights are presented in Table 5.  These 21
mask regions are illustrated in both panels of Figure 7 as the small magneta
circles and illustrate that the masked wavelengths account for a tiny
fraction (3.3\%) of APOGEE pixels.

\begin{deluxetable*}{cc}
\tablecaption{The global mask to remove regions of the spectra not fit well in Arcturus or the Sun by the line-fitting code. This is only an excerpt of the table to show its form and content. The complete table is available in electronic form. \label{tab:global}}
\tablehead{
\colhead{Wavelength} & \colhead{Mask Value}\\
\colhead{(\AA, air)}
}
\startdata
15148.081 & 1.000\\
15148.290 & 1.000\\
15148.499 & 1.000\\
15148.709 & 1.000\\
15148.918 & 1.000\\
15149.127 & 1.000\\
15149.337 & 1.000\\
15149.546 & 1.000\\
15149.755 & 1.000\\
15149.965 & 1.000\\
\nodata & \nodata\\
\enddata
\end{deluxetable*}

\subsection{Astrophysical $\log{(gf)}$ Values}

The differences between the original and final iterated $\log{(gf)}$ values, as $\Delta\log{(gf)}_{\rm (final - initial)}$, are displayed as histograms in Figure 8 for a sample of 5 key chemical species (Fe I, Mg I, Al I, Si I, Ni I).  In the middle panel are shown, as horizontal lines, the maximum amounts that the $\log{(gf)}$ values were allowed to vary depending on the estimated uncertainty in the $gf$-values themselves, using the NIST scale of A, B, C, D; recall that the $\log{(gf)}$ values were allowed to vary by plus-or-minus twice their estimated uncertainty and it is these values that are illustrated by the horizontal lines.  
Overall, there are no significant systematics in the distributions of $\Delta\log{(gf)}$, with all of the elements scattering around zero.  The largest differences are, in general, for the weaker lines that have larger uncertainties in their $gf$-values (or unknown uncertainties) and that are not used in abundance windows in ASPCAP.

As an example of how the weaker atomic lines (which have relatively small impact on derived abundances) have larger uncertainties in their gf-values, Figure 9 is used to illustrate the differences between final and initial $gf$-values for the key species, shown in Figure 8, as a function of the relative line strength ($S=\log{(gf\lambda)} - \theta\chi$).  The species illustrated in Figures 8 and 9 are not special in terms of their gf-values, so the results shown here are, in general similar for other atomic elemental lines.  This particular choice of elements spans a range of nucleosynthetic origins (Fe-peak, $\alpha$-elements, and odd-Z), as well as atomic spectral characteristics. Lines that are used in abundance windows are plotted as the large filled red symbols, while those lines not used as abundance indicators are shown as the smaller open symbols.  Note that for all elements shown, lines that are used to derive abundances are both the stronger lines (clearly detectable, with S$>$-8 or -9, depending on the element in question), and tend to have relatively well-defined $gf$-values (the differences between the final and initial $gf$-values are small).  The weaker lines, with S$<$-9.5, exhibit significantly larger dispersion in their $\Delta$log(gf)-values, were not used to constrain individual abundances, and did not have appreciable effects on abundance fits.  Each panel notes the mean and standard deviations in the $\Delta$log(gf)-values for the lines used in the abundance windows and it is found that all offsets are small ($\le$0.1 dex), suggesting that APOGEE abundance offsets due to atomic line parameters are expected to not be large.

\subsection{Astrophysical Damping Values}

Differences between the original and iterated damping constants for Mg I, Al I, Si I, Fe I, and Ni I are shown in Figure 10; these differences were determined via fits to strong solar lines (not Arcturus).  There are systematic offsets in $\Delta(\log{\Gamma}$) for both Fe I and Ni I, while there are no significant offsets for the Mg I and Si I damping constants.  There is an offset of $\sim-0.1$ dex for Al I, although this is for only three lines, all of which arise from the same lower and upper spectroscopic configurations. 
In this spectral region, the Mg I and Si I lines are stronger than the Fe I and Ni I lines, which may be the reason that the differences between the initial and iterated damping constants display somewhat different behaviors.  The stronger lines (Mg I and Si I) have more well-defined damping wings when compared to the weaker Fe I and Ni I lines in the Sun.

\begin{figure}
\epsscale{0.80}
\plotone{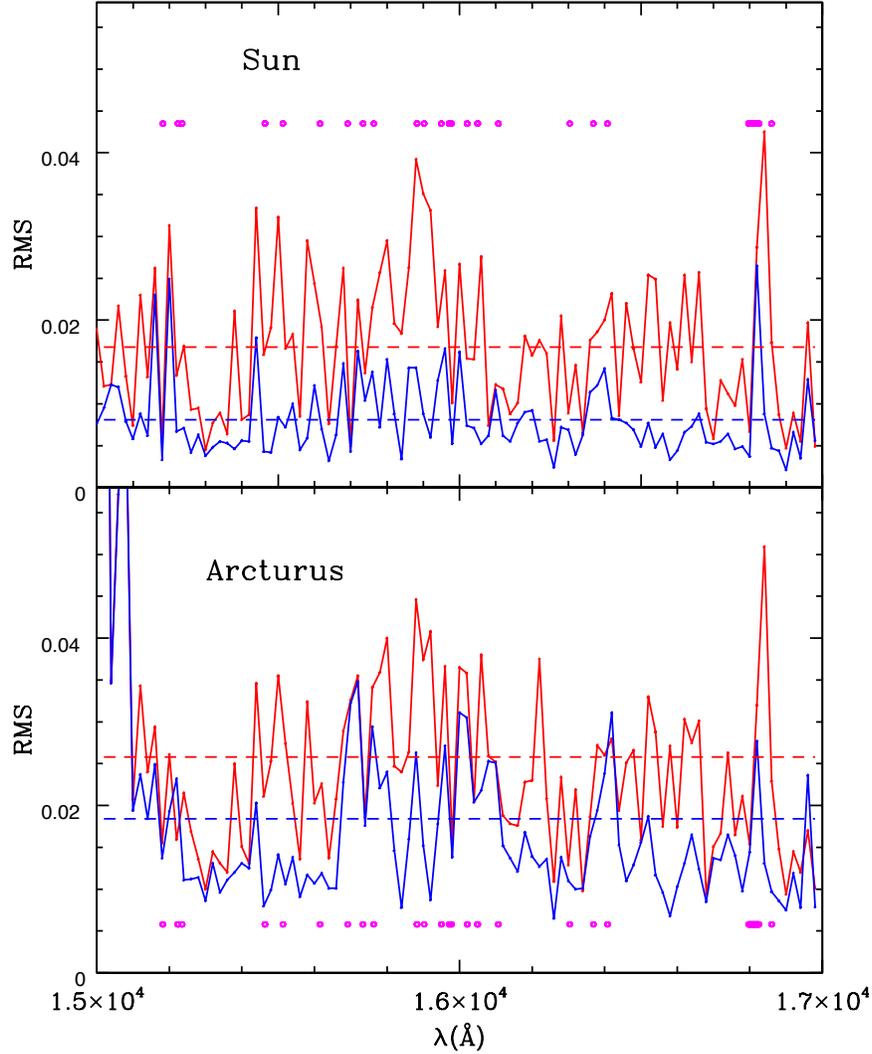}
\caption{The RMS difference between the synthetic and 
observed spectra, calculated in 20\AA~ chunks, with 
the original (dashed line) and astrophysically 
adjusted (solid line) line 
lists for the Sun (upper panel) and Arcturus (lower panel). The dashed lines show the median RMS for the whole spectrum with the adjusted line list.  The small magenta circles in both panels illustrate the masked regions where the final line list produces relative flux mismatches between synthetic and observed spectra of 0.03, or larger, for the Sun, and 0.05, or larger, for Arcturus.  These masked pixels account for 3.3\% of the APOGEE wavelength coverage.
\label{f7}}
\end{figure}

\begin{figure}
\epsscale{0.80}
\plotone{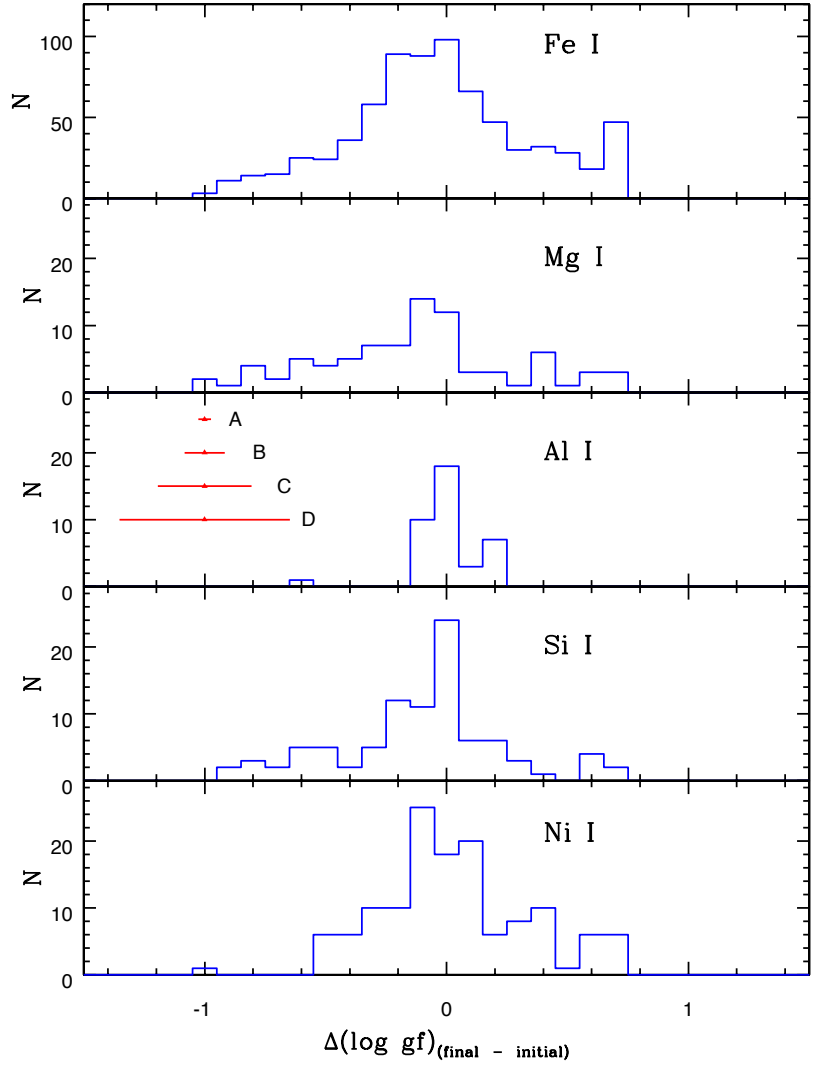}
\caption{Distributions of the differences between the final astrophysically adjusted (iterated) and initial $gf$-values shown for five important species in the APOGEE spectra.  No large systematic differences are found, in general, between the final and initial $gf$-values, with scatter peaked around differences of $\sim$0.0 dex.  The larger differences ($\sim$0.3 dex, or greater) between final and initial $gf$-values are found for weak lines (with no published unceertainties), which were not used for any ASPCAP abundance windows and had negligible affects on the synthetic spectra.
\label{f8}}
\end{figure}

\begin{figure}
\epsscale{0.80}
\includegraphics[width=0.70\textwidth,clip,angle=0]{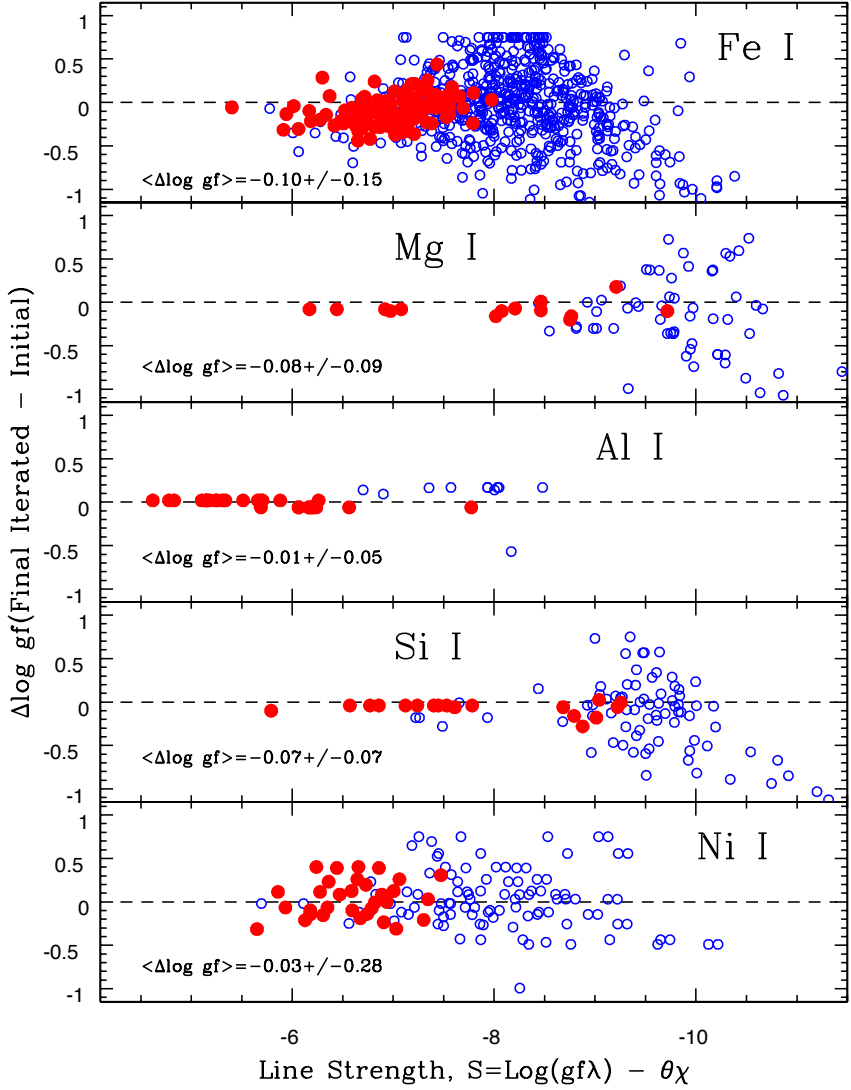}
\caption{Differences between the final and initial log gf values ($\Delta$log gf) for the elemental lines from Figure 8 as a function of relative line strength.  Lines used to measure individual chemical abundances are plotted as filled red symbols, with these transitions representing, mostly, the stronger lines for each species (S$>$-8 or -9), and having well-defined oscillator strengths which result in smaller iterative corrections.  The weaker lines (S$<$-9), with larger uncertainties in their oscillator strengths, have little to no impact on derived abundances.  Means and standard deviations of the $\Delta$log(gf)-values for lines used in abundance windows are shown for each species.
\label{f9}}
\end{figure}

\begin{figure}
\epsscale{0.80}
\plotone{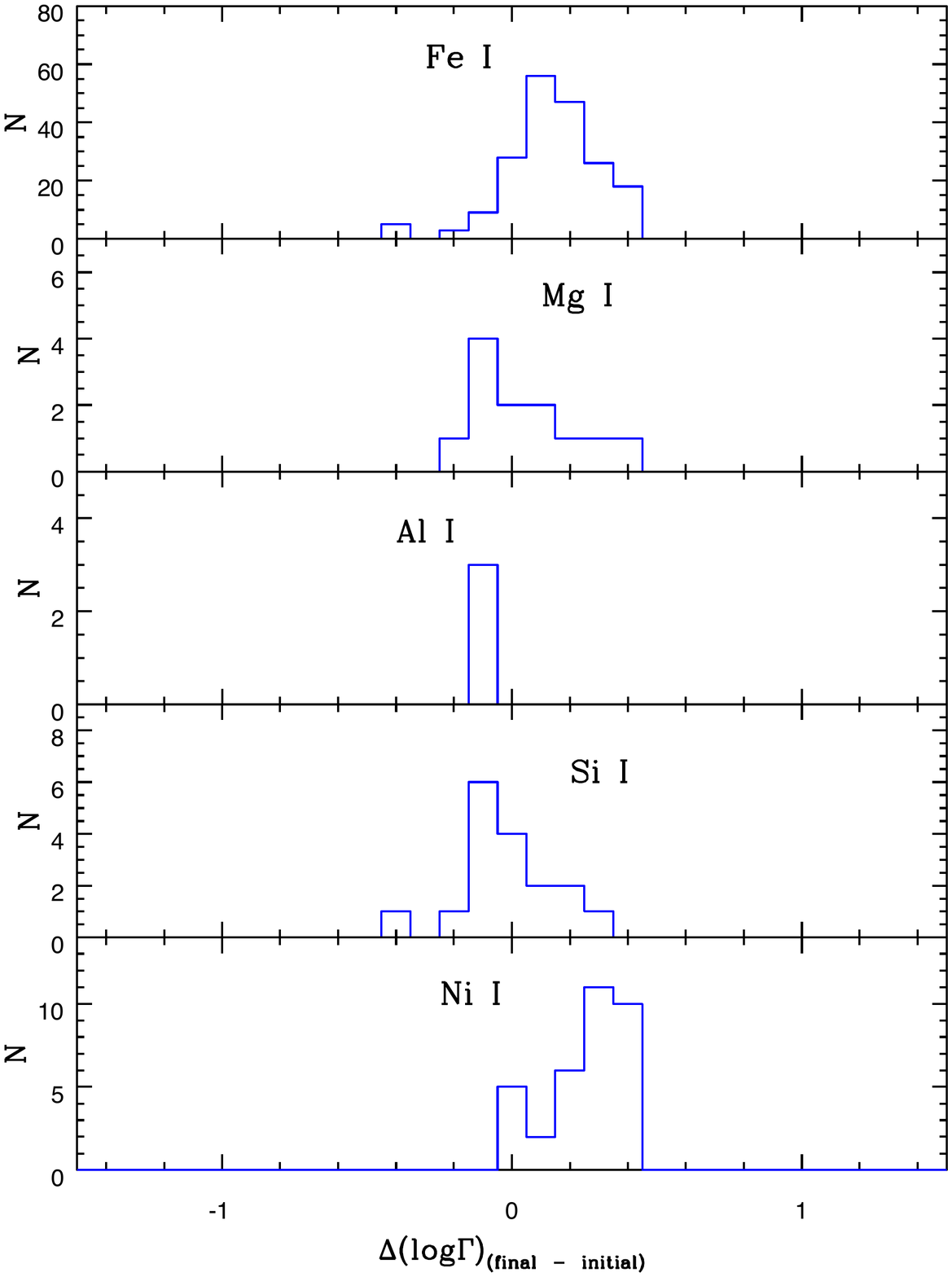}
\caption{The difference between the astrophysically adjusted (new) and original (old) values of damping constants for lines from 5 of the major elemental species.  In the APOGEE spectral window, the Mg I, Al I, and Si I lines are typically stronger than the Fe I and Ni I lines and it is worth noting that the $\Delta$(log$\Gamma$) values for Mg I and Si I scatter about zero (Al I, with only 3 lines, is not considered in this comparison), while the values for the (weaker) Fe I and Ni I lines are offset slightly to positive values.   
\label{f10}}
\end{figure}

\section{Discussion}

\subsection{Missing Lines and Significantly Mismatched Lines}

Even with the efforts made to add spectral lines, syntheses of both Arcturus and the Sun reveal a small number ($\sim$12) of observed, well-defined spectral lines with depths of 5-25\% that are seemingly missing from the DR16 line list.  Not included in this discussion are lines with merely mismatched $gf$-values, as $gf$-values are adjusted (within limits) by fitting Arcturus and the Sun (COD), with weights determined by the relative line depths in the two standard stars; some lines do not fit as well as others in both stars simultaneously, but these lines are clearly in both observed and model spectra.  The true missing lines are presented in Table 5.  The information tabulated includes the measured $\lambda$ (in air), the depth of the line in both Arcturus and the Sun (CoD), along with possible identifications from the DR16 line list, the NIST database, or from Hinkle et al. (1995).  Seven of the missing lines have near-wavelength matches with Fe I lines in the NIST database, with these lines being unclassified.  It is likely that these seven ``matches'' are the unclassified, high-excitation Fe I lines tabulated in NIST.  

\begin{deluxetable} {cccccc}
\tablecaption{Missing Lines from the DR16 Line List and Possible Identifications}

\label{tab2} 
\tablehead{ \colhead{$\lambda$(\AA)$_{air}$} & \colhead{$\alpha$ Boo Depth} &  \colhead{Solar Depth}  & \colhead{Possible ID: DR16 List} & \colhead{Possible ID: NIST} & \colhead{Hinkle et al. ID} 
}
\startdata
15178.23 & 0.20 & 0.23 & V I 78.26 Mn I 78.26 & Fe I 78.25 uncls \\
15402.76 & 0.06 & 0.06 & Fe I $\log{(gf)}=-3.96$ & ... & bl OH (no) \\
15459.40 & 0.15 & 0.17 & ... & Fe I 59.31 uncls & ...  \\
15508.92 & 0.09 & 0.05 & Sc I 508.92 & ...  & ...  \\
15687.44 & 0.14 & 0.12 & ... & ... & ... \\
15729.75 & 0.21 (bl CN) & 0.16 & Fe I 29.81 & Fe I 29.76 & Fe I (probable) \\
15945.26 & 0.16 & 0.10 & Mn I or Ti I ? & Fe I 45.26 uncls & Fe I \\
15967.18 & 0.17 (broad hfs?) & 0.10 (broad hfs?) & Co I $\sim$67.18 (hfs)& ... & CN (no) \\
15973.02 & 0.16 (broad sym) & 0.11 (broad sym) & ... & ... & ... \\
16016.75 & 0.19 & 0.13 & Ni I or Zr I ? & Fe I 16.78 uncls & Fe I (probable) \\
16820.50 & 0.27 & 0.27 & ... & Fe I 20.52 & Fe I (probable) \\
16855.73 & 0.12 & 0.08 & Cr I 55.67 & ... & blend \\
\enddata
\end{deluxetable}

Another set of lines that are not well-fit in either the Sun or Arcturus consist of a small number of, primarily high-excitation, Fe I lines from the Kurucz line list.  In some cases, the model lines are much too strong when compared to the Sun or Arcturus, such that the $gf$-values would need to be decreased by up to 3$-$4 in the log, as well opposite cases where the model lines are orders-of-magnitude weaker than the observed line.  
Kurucz (private communication) notes that the energy level wave functions are linear combinations of basis states for the standard LS-coupling scheme and there can be large uncertainties in the $gf$-values for LS-forbidden lines that can only arise by level mixing in these high excitation energy levels.  Carefully characterizing this sample of Fe I lines would be a useful future exercise.  In addition, the missing broad feature near $\lambda$15967.18\AA\ is likely to be Co I (from the Kurucz line list, $\chi$=5.964 eV), as the hfs components match nicely the shape of the feature, although the $gf$-values would need to be increased by several orders-of-magnitude to fit the feature in both Arcturus and the Sun (presumably the same effect as discussed above for Fe I lines).

In both cases of missing lines, or seriously mismatched lines, the wavelength regions (which fall over very small pieces of the APOGEE wavelength space) are masked from fitting observed spectra to model spectra in ASPCAP (J\"onsson et al. 2020).  All of the missing lines of Table \ref{tab2} are masked out, except the one at $\lambda_{Air}$15402.76\AA.

\subsection{Evaluating the Utility of the DR16 Line List}

The DR16 line list was created from a combination of data based on laboratory measurements, semi-empirical and theoretical calculations from a wide range of sources.  Atomic parameters, most importantly $gf$-values and damping constants were adjusted, within limits defined by the accuracy of these values, by fitting the Sun and Arcturus.  It is worthwhile to now use this line list in a classical analysis of a star and compare the derived parameters and abundances with those results from ASPCAP.  Such an example comparison is presented here.

\begin{figure}
\epsscale{1.20}
\plotone{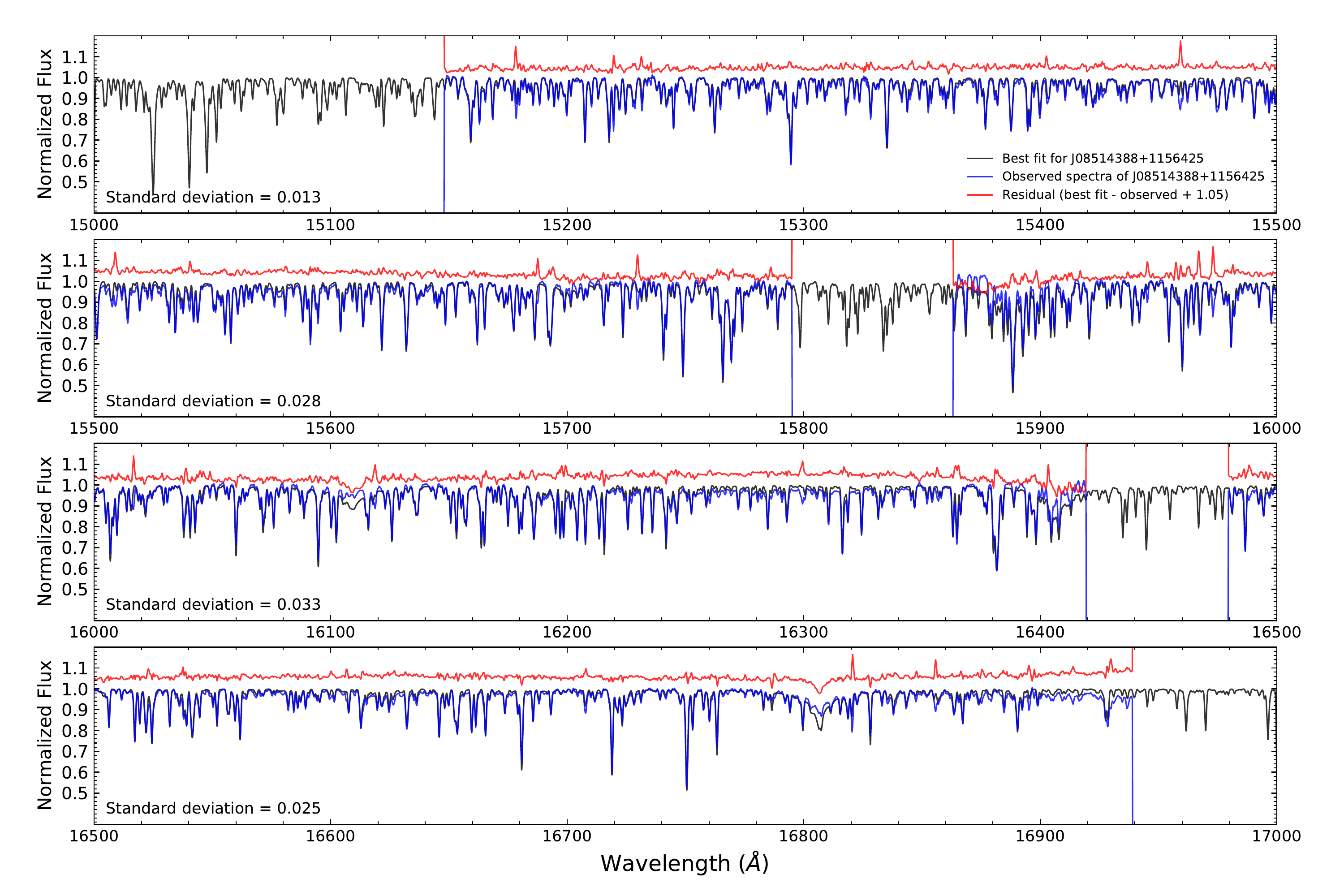}
\caption{Synthetic spectrum computed with the DR16 line list for a red-clump star.  The adopted stellar parameters are from Souto et al. (2018).  The best fit synthetic spectrum is shown in black, the observed in blue, and the difference between them shown by the (shifted) red line.
\label{f11}}
\end{figure}

One relevant example is to apply such a classical analysis to the chemical analysis of a star in a well-studied open cluster, for which the metallicity and age are well constrained. M67 is one such well-studied cluster, having a roughly solar metallicity and solar age. Souto et al. (2018, 2019) analyzed a large number of M67 members from different evolutionary states using APOGEE spectra and found evidence of diffusion in this cluster.

We selected one star member of M67 from Souto et al. (2018) and computed the chemical abundances of several elements. 
The star is 2M08514388+1156425 ($T_{\rm eff}=4820$K, $\log{g} = 2.44$), which is a red clump giant member of M67.
The methodology adopted in the chemical abundance analysis, including the lines measured is discussed in Souto et al. (2019) and the stellar parameters adopted are from Souto et al. (2018). We note that here we are not fitting the entire spectrum as done by ASPCAP. A comparison of the synthetic spectrum with the observed APOGEE spectrum is shown in Figure 9.

There is an overall good fit for most of the spectrum, although in this case the abundances were derived only for selected lines (see Souto et al. for details). The abundances obtained are all within the expected range for a star member of M67, including a recovered iron abundance of $A$(Fe) = 7.48, which is close to the solar abundance adopted in the construction of the line list and adjustment of the $gf$-values.  A comparison of 13 abundances derived classically with ASPCAP values from DR16 finds, from the elements Fe, C, N, O, Na, Mg, Al, Si, K, Ca, Ti, V, and Mn, a mean difference and standard deviation of $\Delta$[X/H]$=+0.04 \pm 0.05$ dex.  This comparison suggests that the DR16 line list can be used as a general tool for quantitative spectroscopy of both red giants and cool main-sequence stars.




\begin{acknowledgments}
VS and KC acknowledge that their work here is supported, in part, by the National Aeronautics and Space Administration under grant 16-XRP16-2-0004, issued through the Astrophysics Division of the Science Mission Directorate, as well as the NSF grant AST-2009507.
TM and DAGH acknowledge support from the State Research Agency (AEI) of the Spanish Ministry of Science, Innovation and Universities (MCIU) and the European Regional Development Fund (FEDER) under grant AYA2017-88254-P.
JH and DB acknowledge support from NSF grant AST-1715898. 
SM has been supported by the J{\'a}nos Bolyai Research Scholarship of the Hungarian Academy of
Sciences, by the Hungarian NKFI Grants K-119517 and GINOP-2.3.2-15-2016-00003 of the Hungarian National Research, Development and Innovation Office, and by the {\'U}NKP-20-5 New National Excellence Program of the Ministry for Innovation and Technology.

We would like to thank the anonymous referee for a careful reading of the initially submitted manuscript and very constructive comments and questions, which improved the final version of this paper.

SDSS-IV is managed by the Astrophysical Research Consortium for the
Participating Institutions of the SDSS Collaboration including the
Brazilian Participation Group, the Carnegie Institution for Science,
Carnegie Mellon University, the Chilean Participation Group, the French
Participation Group, Harvard-Smithsonian Center for Astrophysics,
Instituto de Astrof\'isica de Canarias, The Johns Hopkins University,
Kavli Institute for the Physics and Mathematics of the Universe (IPMU) /
University of Tokyo, Lawrence Berkeley National Laboratory, Leibniz
Institut f\"ur Astrophysik Potsdam (AIP), Max-Planck-Institut f\"ur
Astronomie (MPIA Heidelberg), Max-Planck-Institut f\"ur Astrophysik (MPA
Garching), Max-Planck-Institut f\"ur Extraterrestrische Physik (MPE),
National Astronomical Observatory of China, New Mexico State University,
New York University, University of Notre Dame, Observatrio Nacional /
MCTI, The Ohio State University, Pennsylvania State University, Shanghai
Astronomical Observatory, United Kingdom Participation Group, Universidad
Nacional Aut\'onoma de M\'exico, University of Arizona, University of
Colorado Boulder, University of Oxford, University of Portsmouth,
University of Utah, University of Virginia, University of Washington,
University of Wisconsin, Vanderbilt University, and Yale University.
\end{acknowledgments}


\end{document}